\documentclass[prd, preprint, longbibliography, 11pt]{revtex4-1}

\usepackage{amsmath}
\usepackage{mathtools}
\usepackage{amssymb}
\usepackage{setspace}
\usepackage{graphicx}
\usepackage{natbib}
\usepackage{float}
\usepackage{tensor}
\usepackage[utf8]{inputenc}
\usepackage{amsfonts}
\usepackage{braket}
\usepackage{esint}
\usepackage{breqn}
\usepackage{IEEEtrantools}

\begin{document}
 
 %

\begin{center}
{ \large \bf Relativistic weak quantum gravity}\\
 {\large \bf and its significance for the standard model of particle physics}

\smallskip

\vskip 0.1 in

{\large{\bf Tejinder P.  Singh }}

\smallskip

{\it Tata Institute of Fundamental Research,}
{\it Homi Bhabha Road, Mumbai 400005, India}\\
\smallskip
 {\tt tpsingh@tifr.res.in}

\end{center}

\centerline{\bf ABSTRACT}
\noindent There ought to exist a reformulation of quantum theory, even at energy scales much lower than Planck scale, which does not depend on classical time. Such a formulation is required also for the standard model of particle physics, at the low energies at which it is currently observed. We have proposed such a formulation, by replacing 4D Minkowski spacetime by an octonionic space. Doing so allows us to naturally construct spinor states which describe quarks and leptons having properties as in the standard model.  We conclude that the aforesaid reformulation of quantum theory helps understand why the standard model is what it is. We do not need experiments at ever higher energies to understand the low energy standard model. Instead, we need a better understanding of the quantum nature of spacetime at low energies, such that the quantum spacetime is consistent with the principle of quantum linear superposition. In the present short review article, we give a summary of our ongoing research programme which aims to address these  issues.

\section{Introduction}

\noindent One way in which quantum theory is approximate is through its dependence on an external classical time. The existence of such a time is physically meaningful only when the universe is dominated by classical bodies such as macroscopic objects, e.g. stars and galaxies - their dominance is mutually consistent with a classical spacetime geometry. In their absence, there ought to exist a way to describe quantum dynamics without referring to classical time. We call this latter situation quantum gravitational: it is not the quantization of gravitation, but  an essential reformulation of quantum theory without recourse to a classical spacetime geometry. Such a situation can arise in principle, as noted by Penrose \cite{Penrose:96}, when a massive object is in a quantum superposition of two different position states, because the definition of time at any chosen point in space becomes ambiguous. Nonetheless we should have a theory for describing the dynamics of such a superposition. This is an example of low energy non-relativistic quantum gravity, as opposed to quantum gravity at the Planck energy scale. The corresponding relativistic low-energy quantum gravitational case arises if we wish to describe the quantum theory of the standard model of particle physics at low energies, without any reference to a background Minkowski space-time. In particular, it seems natural that fermions in quantum theory should be defined on a [non-commutative] spinor space-time, whereas their description on Minkowski space-time can only be an approximate one. As we explain below, doing so restricts the allowed symmetries and properties of the fermions, such as the ratios of their electric charges and their masses, reproducing precisely the observed symmetries of the standard model of particle physics. In this way, not only does space-time tell matter how to move, but it also tells matter what to be! We do not need new physics at high energies in order to understand the standard model at low energies. Instead, we need to properly understand the quantum structure of spacetime at low energies, when the sources are fermions in quantum superposition of different position states.

The UV aspect of quantum gravity, and its IR aspect, can both be jointly characterised through a common set of conditions. To begin with, a phenomenon will be said to be quantum gravitational in nature, if one or more of the following three conditions are satisfied: (i) the resolution of the time scale of interest is of the order of Planck time $t_P$,
(ii) the resolution of the length scale of interest is of the order of Planck length $L_P$,
(iii) the action of all subsystems of interest is of the order of Planck's constant $\hbar$. While the first two conditions are obvious, the third one could do with some explanation, because usually a system with action of order $\hbar$ is described as being quantum, but not quantum gravitational. However, this is because a dominant classical space-time background is already assumed to be present, and as we noted above, such a background classical gravitation is sourced by macroscopic classical bodies which all have action much greater than $\hbar$. Hence, a quantum system in the usual formulation of quantum theory is not quantum gravitational, whereas if all the sub-systems have action comparable to $\hbar$, it means no macroscopic bodies are present, and hence the situation is quantum gravitational. [By a sub-system we mean a set of degrees of freedom which are all entangled with each other, but not with those of another sub-system. For example, in a system such as a beam of fullerene molecules, each molecule of fullerene is a quantum subsystem; the spacetime produced by such a beam is nonclassical, and we would like to have a mathematical description for such a nonclassical geometry].

In other words, even if only (iii) is satisfied, quantum superpositions are necessarily present, and hence there is no classical spacetime. That is why this is a quantum gravitational situation. If (iii) as well as either of  (i) or (ii) hold then this is a Planck energy scale phenomenon. If only (iii) holds but neither (i) nor (ii) hold then we have low energy quantum gravity [this could be relativistic or non-relativistic]. Space-time is then non-classical even at low energies, and this non-classicality of spacetime fixes the standard model. 
In particular, the low energy fine structure constant $e^2/\hbar c \equiv e^2 t_P/\hbar L_P \sim 1/137$ is order one in these quantum gravitational units [$L_P$, $t_P$, $\hbar$]. In other words, the square of the electric charge of the electron,  being 1/137 and hence order unity in these units, is determined as a low-energy quantum gravitational effect. This is borne out in our work \cite{GRFEssay2021}. We construct an action principle in which the only free parameters are Planck length, Planck time and Planck's constant. Every other physical constant, dimensional as well as dimensionless, is intended to be derived in terms of these three parameters. Note that in comparison with the usual studies of quantum gravity, we have traded Planck mass for Planck's constant $\hbar$. This enables us to treat both UV and IR quantum gravity in the same analysis. Constants such as Newton's gravitational constant $G$, speed of light $c$, and  Planck mass $m_P$ are all now regarded as derivative quantities.

To construct the aforesaid desired reformulation of quantum theory, we must propose a new physical space [which replaces 4D Minkowski space-time] on which fermionic spinors are to be defined. We must then propose a dynamics on this new physical space, and work out the consequences of the new space and the new dynamics, and compare them with known but unexplained data. Finally we must show how classical spacetime geometry, and quantum (field) theory as we know it, are recovered from this new formulation. Below, we outline the current status of this programme, pointing out what has been achieved so far, and what is still work in progress and remains to be addressed / completed. A  more detailed report is given in \cite{Singhreview}.

\section{Division algebras and Clifford algebras}

To construct the new physical space, we consider number systems more general than real numbers [the latter of course label the Lorentzian $R^4$ space-time manifold]  and which retain the mathematical property of division, apart from addition, subtraction, and multiplication. There are only four such normed division algebras: the reals $\mathbb R$, the complex numbers $\mathbb C$, the quaternions $\mathbb H$ and the octonions $\mathbb O$. The quaternion algebra is non-commutative, whereas the octonion algebra is non-commutative as well as non-associative. A quaternion is the four-dimensional object $(a_0 + a_1 \hat i +a_2 \hat j + a_3 \hat k)$ where the $a_i$ are real numbers, and $\hat i, \hat j, \hat k$ are unit imaginaries obeying the following product rules
\begin{equation}
{\hat i}^2 = {\hat j}^2 = {\hat k}^2 = -1\; , \qquad \hat i \hat j = - \hat j \hat i = \hat k\; , \qquad \hat j \hat k = - \hat k \hat j = \hat i \; , \qquad \hat k \hat i = - \hat i \hat k = \hat j
\label{qua1}
\end{equation}
An octonion is the eight-dimensional mathematical object 
\begin{equation}
O = a_0 + a_1 e_1 + a_2 e_2 + a_3 e_3 + a_4 e_4 + a_5 e_5 + a_6 e_6 + a_7 e_7
\end{equation}
where the $a_i$ are real numbers, and the $e_i$ are seven imaginary units [$e_i^2 = - 1$] which anti-commute and satisfy the following product rules given by the Fano plane shown in Fig. 1 below.
\begin{figure}[h]
\centering
\includegraphics[width=7.5cm]{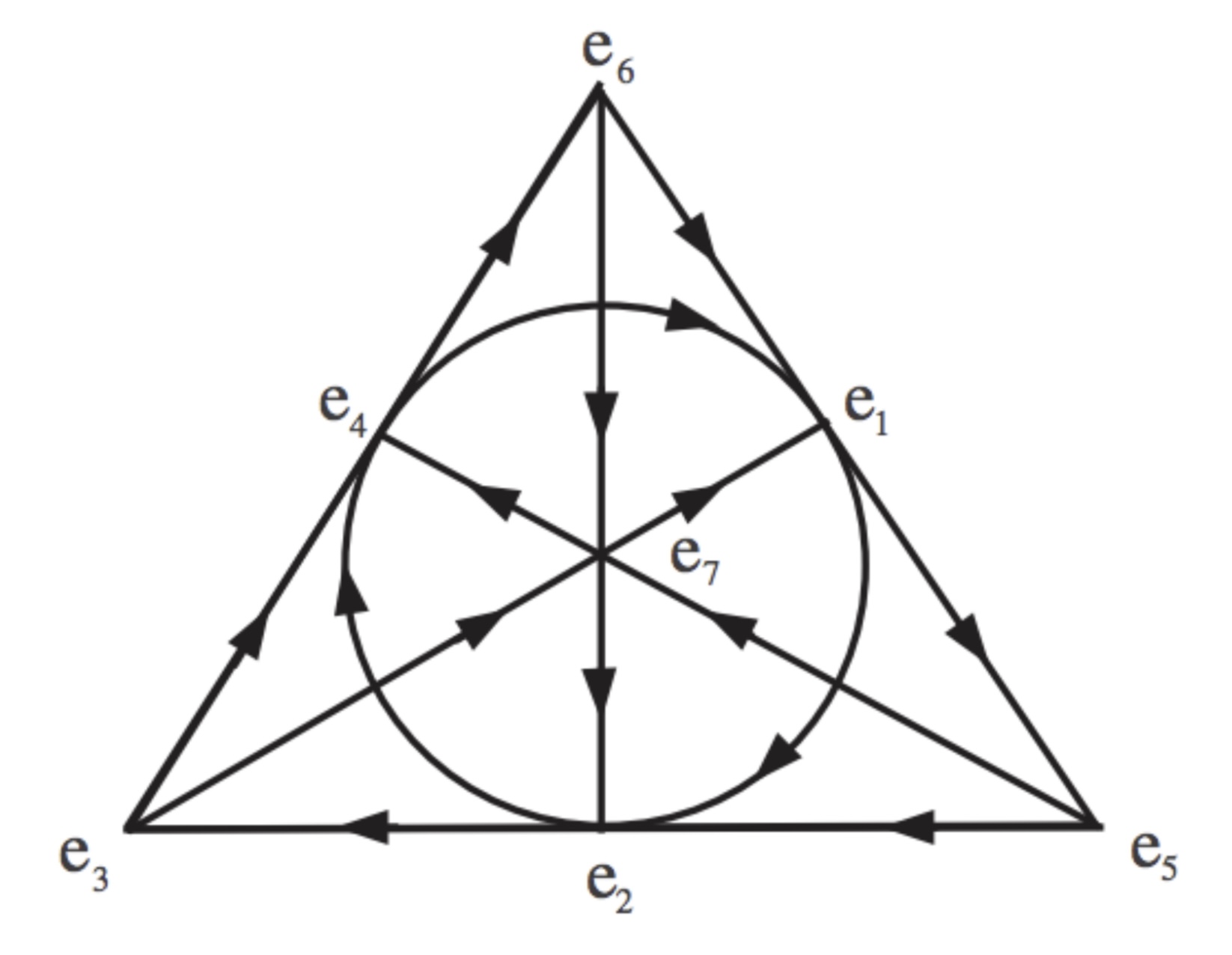}
\caption{The Fano plane}
\end{figure}
There are seven quaternionic subsets in the Fano plane, given by the three sides of the triangle, the three altitudes, and the incircle. Multiplication of points lying along a quaternionic subset in cyclic order (as per the arrow) is given by $e_ie_j = e_k$, whereas $e_je_i = -e_k$. The automorphisms of the quaternions form the group $SO(3)$, the group of rotations in three dimensions, which of course is what Hamilton invented the quaternions for. The automorphism group of the octonions is $G_2$, the smallest of the five exceptional Lie groups $G_2, F_4, E_6, E_7$ and $E_8$. The group $F_4$ is the automorphism group of the exceptional Jordan algebra $J_3(8)$ and $E_6$ of the complexified exceptional Jordan algebra. Of the five exceptional groups, only $E_6$ has complex representations \cite{Ramond1976}.

Complex quaternions and complex octonions, especially the latter, have been seriously investigated for their role in understanding the properties of quarks and leptons in the standard model of particle physics \cite{Dixon, Gursey, f1, f2, f3, Chisholm, Trayling, Dubois_Violette_2016, Todorov:2019hlc, Dubois-Violette:2018wgs, Todorov:2018yvi, Todorov:2020zae, ablamoowicz, baez2001octonions, Baez_2011,  baez2009algebra, f1, f2, f3, Perelman,  Gillard2019, Stoica, Yokota, Dray1, Dray2, lisi2007exceptionally, Ramond1976}.  The automorphism groups of the octonions and of the exceptional Jordan algebra show evidence that they contain within themselves the symmetry groups of the standard model. Clifford algebras made from the quaternions and the octonions are used to construct spinors and these have associated properties which match those of quarks and leptons. Furthermore, the exceptional group $E_6$ shows evidence that it includes within itself Lorentz symmetry, which can be unified with the internal symmetries of the standard model, and coupled with an appropriate dynamics, lead us to the formulation of quantum theory without classical time. Thus our programme differs from earlier works in that division algebras are being used to define the structure of quantum spacetime, in addition to being used to explore the standard model \cite{Singhreview}.

We recall that a Clifford algebra over the field $\mathbb{R}$ is an associative algebra written as $Cl(p,q)$ such that there is a $n = p + q$ dimensional vector space $V = \{e_1, e_2, ..., e_n\} $ satisfying:
\begin{equation}
    \{e_i, e_j\} \equiv e_ie_j + e_je_i = 2\eta_{ij}\mathbb{I} 
\end{equation}
Here the vector space $V$ is called the generating space of $Cl(p,q)$ and $\eta_{ij} = 0$ if $i \neq j$, $\eta_{ii} = 1$ for $i = 1,..., p$, $\eta_{ij} = -1$ for $i = p+1,...., p+q$. There is an identity vector in the algebra, such that $a.1 = 1.a = a$, $\forall a \in Cl(p,q)$. The elements of a Clifford algebra can be constructed by multiplication of generating vectors, therefore, it can be be checked that the dimension of the Clifford algebra $Cl(p,q)$ will be $\sum_{i=0}^{n} {}^nC_i = 2^n$.

As an example,  the algebra $Cl(0,0)$  will correspond to the real numbers $\mathbb{R}$. There are no generating vectors, just the trivial identity vector in this case. $Cl(0,1)$ will correspond to the complex numbers, here we have only one generating vector \say{$i$} with $i^2 = -1$. For $Cl(0,2)$ we need two generating vectors $e_1^2 = -1$ and $e_2^2 = -1$. The algebra will then be $Cl(0,2)$ = $\{1, e_1, e_2, e_1e_2\}$, we identify this algebra with the division algebra of quaternions $\mathbb{H}$. Therefore from the first three Clifford algebras we obtain the division algebras of $\mathbb{R}, \mathbb{C}$, and $\mathbb{H}$. The octonions $\mathbb{O}$ do not make Clifford algebras naturally because of their non-associativity but they can be used to make the associative algebra $Cl(6)$ by introducing the octonionic chains, as was shown by Furey \cite{f1}.

The next Clifford algebra in this series is $Cl(0,3)$, the generating vectors will be $e_1^2 = -1, e_2^2 = -1$, and $e_3^2 = -1$. The Clifford algebra $Cl(0,3)$ will have eight elements:\\  $\{1, e_1, e_2, e_3, e_1e_2, e_2e_3, e_3e_1, e_1e_2e_3\}$. We can break this algebra into sum of two quaternionic parts:
\begin{equation}
(1, e_1, e_2, e_1e_2), \quad(e_1e_2e_3, e_2e_3, e_3e_1, e_3)
\end{equation}
It is important to note that $e_3 \neq e_1e_2$ as is the case for quaternions. We can identify the left set as the quaternions; in the right set $(e_1e_2e_3)^2 = 1$ whereas the square of the other three elements is $-1$. If we call $e_1e_2e_3$ as $\omega$ then the right set becomes $\omega(1, -e_1, -e_2, -e_1e_2)$. Thus the right set is $\omega$ times the quaternions. $\omega$ here is a split complex number; analogous to $i$ which squares to -1, the split complex numbers square to 1 but are neither 1 nor $-1$. The conjugate of $\omega$ is $-\omega$. Thus, the algebra $Cl(0,3)$ is known as the split-biquaternions. It can be written as $\mathbb{D} \otimes \mathbb{H}$, where $\mathbb{D} \equiv (1, \omega)$. Notice that $\mathbb{D} \otimes \mathbb{H} \cong \mathbb{H} \oplus \mathbb{H}$.

Next we consider Clifford algebras on the complex field. The Clifford algebra $Cl(n)$ is defined on the complex field by an $n$ dimensional vector space $V = \{e_1,....,e_n\}$ such that:
\begin{align}
    \{e_i, e_j\} = 0,\quad i \neq j\\
    e_i^2 = 1.
\end{align}
It is interesting to note that the Clifford algebra $Cl(n)$ on the complex field can be obtained from the Clifford algebra $Cl(p,n-p)$ on the real field by the following relation: $Cl(n) = \mathbb{C} \otimes Cl(p,n-p)$ where $0 \leq p \leq n$.
Therefore, we note that we can get the algebra $Cl(3)$ from complexification of Cl(0,3) and the final algebra would be complex split-biquaternions $\mathbb{C} \otimes (\mathbb{H} \oplus \omega\mathbb{H})$. The Clifford algebra $Cl(0,7)$ is the algebra of split $8 \times 8$ real matrices, $\mathbb{R}[8] \oplus \omega\mathbb{R}[8]$. On complexifying we will get $Cl(7)$ which is $\mathbb{C}[8] \oplus \omega\mathbb{C}[8]$. It has been shown by Furey in \cite{f1} that $Cl(6) = \mathbb{C}[8] \cong \overleftarrow{\mathbb{C}} \otimes \overleftarrow{\mathbb{O}}$. Here $\overleftarrow{\mathbb{C}} \otimes \overleftarrow{\mathbb{O}}$ is the algebra of complex octonionic chains, defined as a series of maps acting on a function $f \in \mathbb{C}\otimes\mathbb{O}$ from left to right. Therefore $Cl(7) \cong \overleftarrow{\mathbb{C}} \otimes \overleftarrow{\mathbb{O}} + \omega\overleftarrow{\mathbb{C}} \otimes \overleftarrow{\mathbb{O}}$, the complex split bioctonions.

\section{Standard model symmetries from the algebra of complex octonions}

Consider now the complex quaternions $\mathbb C \times \mathbb H$ and the Clifford algebra $Cl(2)$ which can be constructed from them. The generating vector space of $Cl(2)$  is $W = \{i\hat i, i\hat j\}$ and the third direction $\hat k$ is kept fixed (notation as in (\ref{qua1})). A subspace $U \subset W$ is said to be a maximal totally isotropic subspace (MTIS) of $W$, if:
\begin{equation}
    \{\alpha_i, \alpha_j\} = 0 \quad \forall \alpha_i \in U 
\end{equation}
We can see that the MTIS for $Cl(2)$ will be one dimensional with either the element $\alpha = \frac{\hat i + i\hat j}{2}$ or $\frac{-\hat i + i\hat j}{2}$. The primitive idempotent $V$ of this algebra can be defined as $\alpha \alpha^{\dagger}$. We can now construct spinors as minimal left ideals of $Cl(2)$ by left multiplication of $\mathbb{C} \otimes \mathbb{H}$ on the idempotent. The left ideals thus obtained will be Weyl spinors of distinct chirality and spin.
If we use $\alpha = \frac{\hat i + i\hat j}{2}$ to make the idempotent the resulting Weyl spinor will be:
\begin{equation} \label{1}
\psi_R = \epsilon^{\downarrow \uparrow}\alpha^{\dagger}V + \epsilon^{\uparrow \uparrow}V
\end{equation}
where $\epsilon^{\downarrow \uparrow}, \epsilon^{\uparrow \uparrow} \in \mathbb{C}$, $V$ is the idempotent. This right-handed Weyl spinor $\psi_R$ can be interpreted as an entangled state of $V$ and $\alpha^{\dagger}V$, where $V$ and $\alpha^{\dagger}V$ are two different leptons.
The difference between these two leptons is evident from the number operator 
\begin{equation}
    N = \alpha^{\dagger}\alpha
\end{equation}
The spinors $V$ and $\alpha^{\dagger}V$ are eigenvectors of this Hermitian number operator with eigenvalues 0 and 1 respectively. These eigenvalues correspond to the charge of the leptons. Therefore $V$ will be interpreted as the neutrino and $\alpha^{\dagger}V$ will be interpreted as the charged lepton, and the $U(1)$ number operator is identified with $U(1)_{em}$.

We can similarly obtain a left handed Weyl spinor $\psi_L$ if we use the MTIS $\alpha = \frac{-\hat i + i\hat j}{2}$. The idempotent obtained now will be $V^*$. Instead of $\alpha^{\dagger}$, we will now use $\alpha$ to obtain the excited state from the idempotent. The obtained left-handed Weyl spinor is
\begin{equation} \label{2}
\psi_L = \epsilon^{\uparrow \downarrow} \alpha V^* + \epsilon^{{\downarrow \downarrow}}V^*
\end{equation}
where $\epsilon^{\uparrow \downarrow}, \epsilon^{\downarrow \downarrow} \in \mathbf{C}$. Individually we can identify $V^*$ and $\alpha V^*$ as a neutrino and a charged lepton. The left handed leptons will be the anti-particles of right-handed leptons and will be related through complex conjugation $*$.

The SL(2,C) symmetry responsible for Lorentz invariance of vectors and spinors is already present in the Cl(2) algebra. We can write a 4-vector in Minkowski spacetime using the quaternions as follows:
\begin{equation} \label{3}
    V = v_0 + v_1\hat i + v_2\hat j + v_3\hat i \hat j
\end{equation}
Any number $s$ in the $\mathbb{C} \otimes \mathbb{H}$ algebra which does not have a component along the real direction can be shown to be the generator of the Lorentz algebra. For $s = s_1\hat i + s_2i\hat i + s_3\hat j + s_4i\hat j + s_5\hat i\hat j + s_6i\hat i \hat j$, the Lorentz operator $e^{is}$ will bring about boosts and rotations. Notice that $i\hat i, i\hat j, i\hat i \hat j$ are the Pauli matrices and the subgroup SU(2) is already present in SL(2,C). This SU(2) is the usual spin group responsible for rotational symmetry in 3-D space. The complex quaternions enable the transition from $SO(3)$ to $SO(1,3) \sim SL(2,C)$.

The complex quaternions provide the simplest example of a division algebra being used to construct spinors defining fermions in the standard model, as well as internal as well as spacetime symmetries, in this case these are respectively one generation of leptons, the electromagnetic symmetry, and 4D Lorentz symmetry. Only two values of electric charge are possible: 0 and 1. 

In an analogous manner, this construction can be repeated with the complex octonions $\mathbb C \times \mathbb O$. Keeping one out of the seven imaginary directions fixed yields a three-dimensional MTIS and the Clifford algebra $Cl(6)$. The automorphism group $G_2$ of the octonions has two maximal subgroups: $U(3)\sim SU(3)\times U(1)$, and $SO(4)\sim SU(2)\times SU(2)$, and these have an intersection $SU(2)\times U(1)$. The first of these two sub-groups is the element preserver group of the octonions (i.e. one imaginary direction remains fixed), whereas the second one is the stabiliser group of the quaternions in the octonions.  The MTIS is employed to generate an eight dimensional basis of spinor states, by acting on the idempotent with $\mathbb C \times \mathbb O$, and analogous to the quaternionic case above, a $U(1)$ number operator is made from the MTIS generators. Two of these states are singlets under the $SU(3)$ and respectively have number operator eigenvalues 0 and 3. Three states are anti-triplets under $SU(3)$ and have number operator eigenvalue 1 while the remaining three states are triplets under $SU(3)$ and have number operator eigenvalue 2. These non-trivial properties permit the inference that one-third of the number operator can be identified with the electric charge, and the $U(1)$ with $U(1)_{em}$. The singlet states are therefore the neutrino and the positron, whereas the anti-triplet states are the anti-down quarks with charge 1/3, and the triplet states are the up quarks with electric charge 2/3. The $SU(3)$ symmetry is identified with the $SU(3)_{color}$ of QCD, and the $U(1)$ with $U(1)_{em}$ of electrodynamics. In this way the unbroken $SU(3)_c \times U(1))_{em}$ symmetry of quarks and leptons is revealed to be a consequence of the algebra of the octonions. These important results were shown by Furey \cite{Fureycharge}.

The MTIS for the above-mentioned $Cl(6)$ will be 3-dimensional with the following elements:
\begin{equation}
    \alpha_1 = \frac{-e_5 + ie_4}{2}, \quad \alpha_2 = \frac{-e_3 + ie_1}{2}, \quad \alpha_3 = \frac{-e_6 + ie_2}{2}
\end{equation}
After defining $\Omega = \alpha_1\alpha_2\alpha_3$, the idempotent will be $\Omega\Omega^{\dagger} = \alpha_1\alpha_2\alpha_3\alpha_3^{\dagger}\alpha_2^{\dagger}\alpha_1^{\dagger}$. On left-multiplying the idempotent with elements of MTIS we obtain the following desired excited states:
\begin{align}
    \overline{\mathcal{V}} = \Omega\Omega^{\dagger} = \frac{ie_7 + 1}{2}\\
    V_{ad1} = \alpha_{1}^{\dagger}\mathcal{V} = \frac{e_5 + ie_4}{2}\\
    V_{ad2} = \alpha_{2}^{\dagger}\mathcal{V} = \frac{e_3 + ie_1}{2}\\
    V_{ad3} = \alpha_{3}^{\dagger}\mathcal{V} = \frac{e_6 + ie_2}{2}\\
    V_{u1} = \alpha_{3}^{\dagger}\alpha_{2}^{\dagger}\mathcal{V} = \frac{e_4 + ie_5}{2}\\
    V_{u2} = \alpha_{1}^{\dagger}\alpha_{3}^{\dagger}\mathcal{V} = \frac{e_1 + ie_3}{2}\\
    V_{u3} = \alpha_{2}^{\dagger}\alpha_{1}^{\dagger}\mathcal{V} = \frac{e_2 + ie_6}{2}\\
    V_{e+} = \alpha_{3}^{\dagger}\alpha_{2}^{\dagger}\alpha_{1}^{\dagger}\mathcal{V} = -\frac{(i + e_7)}{2}
\end{align}
The following generator for $U(1)_{em}$ provides charge to the quarks and leptons
\begin{equation}
    Q = \frac{\alpha_1^{\dagger}\alpha_1 + \alpha_2^{\dagger}\alpha_2 + \alpha_3^{\dagger}\alpha_3}{3}
\end{equation}
Therefore we get one generation of quarks and leptons from $Cl(6)$; the anti-particles will be related through complex conjugation $*$. As can be seen, each quark comes in three colours. It has been shown by authors \cite{f1, Gillard, f2, f3} that the algebra ${\mathbb{C}} \otimes {\mathbb{O}}$ already has the symmetry group $SU(3)$ present in it. The $SU(3)$ generators in terms of octonions are given by:
\begin{align}
    \Lambda_1 &= -\alpha_2^{\dagger}\alpha_1 - \alpha_1^{\dagger}\alpha_2 & \Lambda_5 &= -i\alpha_1^{\dagger}\alpha_3 + i\alpha_3^{\dagger}\alpha_1 \\
    \Lambda_2 &= i\alpha_2^{\dagger}\alpha_1 - i\alpha_1^{\dagger}\alpha_2 & \Lambda_6 &= \alpha_3^{\dagger}\alpha_2 - \alpha_2^{\dagger}\alpha_3 \\
    \Lambda_3 &= \alpha_2^{\dagger}\alpha_2 - \alpha_1^{\dagger}\alpha_1 & \Lambda_7 &= i\alpha_3^{\dagger}\alpha_2 - i\alpha_2^{\dagger}\alpha_3 \\
    \Lambda_4 &= -\alpha_1^{\dagger}\alpha_3 - \alpha_3^{\dagger}\alpha_1 & \Lambda_8 &= -\frac{(\alpha_1^{\dagger}\alpha_1 + \alpha_2^{\dagger}\alpha_2 - 2\alpha_3^{\dagger}\alpha_3)}{\sqrt(3)}
\end{align}
It can be checked using the charge generator that $\Omega$ has a charge $-1$, whereas $\Omega^{\dagger}$ has a charge 1. Right multiplication of $\Omega$ on the particles  changes their isospin from up to down, whereas the right multiplication of $\Omega^{\dagger}$ on the anti-particles changes their isospin from down to up. Therefore, $\Omega$ mimics the $W^-$ boson and $\Omega^{\dagger}$ mimics the $W^+$ boson. This is another hint  in this approach to the derivation of the standard model, that the $SU(2)_L$ weak symmetry, along with $U(1)_{em}$, is derivable from $SU(3)_{color}$, provided we also involve the other $SU(2)$ of $G_2$ which is not in the intersection of the two maximal subgroups of $G_2$. Recall that we are now working with complex octonions, therefore this $SU(2)$ is extended to an $SL(2,C)$ which is related to the Clifford algebra $Cl(2)$ as discussed above. Below we return to discuss the electroweak symmetry in the language of the octonion algebra.

\section{Complex bioctonions and  left-right symmetric extension of the standard model}

Note that in the construction of $Cl(2)$ from the complex quaternions, and $Cl(6)$ from the complex octonions, one imaginary direction was kept fixed. What happens if this condition is dropped? Doing so  leads us to complex biquaternions and the algebra $Cl(3)$, and to the complex biquaternions and the algebra $Cl(7)$, and therefrom to an extension of the standard model to the Left-Right symmetric model including sterile neutrinos. We will argue that this right-handed extension represents `would-be-gravity' from which space-time geometry and gravitation as we know it are emergent. Furthermore, just as we inferred quantisation of electric charge from the color-electro symmetry above, the L-R extension of the standard model permits us to deduce the observed mass ratios for quarks and charged leptons, from the eigenvalues of the exceptional Jordan algebra $J_3(8)$ \cite{vvs}. The symmetry group is extended from $G_2$ to $F_4$ and $E_6$, and we will subsequently examine the subgroups of $E_6$, and their intersection.

We saw that $Cl(3)$ algebra can be obtained by complexification of $Cl(0,3)$ algebra. We also saw that $Cl(0,3)$ is the spilt biquaternions, therefore we will call $Cl(3)$ as complex split biquaternions $\mathbb{C} \otimes \mathbb{D} \otimes \mathbb{H}$. Recall that we wrote $Cl(0,3)$ as $\mathbb{D} \otimes \mathbb{H} $, this is isomorphic to $\mathbb{H} \oplus \mathbb{H}$, and the $\omega$ is usually left out. However, there possibly is  physical importance for $\omega$ : the symmetry laws will be same for $\mathbb{H}$ and $\omega \mathbb{H}$ (which we call as the omega space), because the multiplication in any Lie algebra has two terms and the $\omega$ will get squared to one. Despite the fact that the two spaces will have similar symmetry laws, the two spaces are different in terms of chirality and $\omega$ can play a crucial role in understanding the constant interaction of left-right fermions with the Higgs.

We wrote the Cl(0,3) algebra as a sum of two sets:
\begin{equation}
\quad (1, e_1, e_2, e_1e_2), \quad \omega(1, -e_1, -e_2, -e_1e_2)
\end{equation}
It is important to note that the two sets have opposite parity. If we create leptons from the two complex quaternions in $Cl(3)$ we will get two sets of leptons with opposite chirality.
The MTIS for the left set of complex quaternions will be either $\alpha = \frac{e_1 + ie_2}{2}$ or $\frac{-e_1 + ie_2}{2}$. The particles and anti-particles created will be
\begin{align}
    \overline{\mathcal{V}}_{R} &= \frac{1 + ie_1e_2}{2} & \mathcal{V}_{L} &= \frac{1 - ie_1e_2}{2}\\
    e^{+}_{R} &= \frac{-e_1 + ie_2}{2} &
    e^{-}_{L} &= \frac{-e_1 - ie_2}{2} 
\end{align}
This way we have created the left-handed neutrino $\mathcal{V}_{L}$ and electron $e^{-}_{L}$ along with their right-handed anti-particles; the anti-neutrino $\overline{\mathcal{V}}_{R}$ and positron $e^{+}_{R}$. The particles and anti-particles are related to each other through complex conjugation $*$.

Similarly we can get leptons from the right set of complex quaternions. The MTIS will be either $\alpha = \omega(\frac{-e_1 - ie_2}{2})$ or $\omega(\frac{e_1 - ie_2}{2})$. The particles and anti-particles created will be
\begin{align}
    \overline{\mathcal{V}}_{L} &= \frac{1 + ie_1e_2}{2} & \mathcal{V}_{R} &= \frac{1 - ie_1e_2}{2} 
   &  e^{+}_{L} &= \omega\frac{(e_1 - ie_2)}{2} &
    e^{-}_{R} &= \omega\frac{(e_1 + ie_2)}{2} 
\end{align}
As stated before the particles created from the right set of complex quaternions have opposite chirality to the particles created from left set of complex quaternions. Therefore, we now have the right-handed neutrino $\mathcal{V}_{R}$ and the right-handed electron $e^{-}_{R}$, along with their left-handed anti-particles; the anti-neutrino $\overline{\mathcal{V}}_{L}$ and positron $e^{+}_{L}$. It has been pointed in \cite{f1} that parity transformation can be brought by $e_i \rightarrow -e_i$, our result here is consistent with this fact.
The generator for $U(1)_{em}$ i.e.
$
    Q = \alpha^{\dagger}\alpha
$
is present in $Cl(3)$ and provides charge to the leptons of both the sectors.

Next, using $Cl(6)$ we can create one generation of fermions with distinct chirality. Left-handed particles from $Cl(6)$ will have right-handed anti-particles. We will now investigate $Cl(7)$ to see that it can be naturally written as a sum of two $\overleftarrow{\mathbb{C}} \otimes \overleftarrow{\mathbb{O}}$ with opposite parity. Throughout our analysis octonions should be treated as octonionic chains acting on the function $f = 1$; it is important to note that unlike in $Cl(6)$ $\overleftarrow{e_7} \neq \overleftarrow{e_1e_2e_3e_4e_5e_6}$. For our convenience we replace $\overleftarrow{e_1e_2e_3e_4e_5e_6}$ by $\overleftarrow{e_8}$. Cl(0,7) can be made from two sets of octonions:
\begin{equation}
    (1, e_1, e_2, e_3, e_4, e_5, e_6, e_8) \oplus \omega(1, -e_1, -e_2, -e_3, -e_4, -e_5, -e_6, -e_8) 
\end{equation}
Here $\omega = \overleftarrow{e_1e_2e_3e_4e_5e_6e_7}$. The above line of reasoning is not direct to see, but we can understand it in the following way. $Cl(0,7)$ is split $8\times 8$ real matrices $R[8]\oplus R[8]$, therefore $Cl(7)$ will be $C[8]\oplus C[8]$, the Clifford algebra $Cl(6)$ is also $C[8]$ and is isomorphic to complex octonionic chains. Therefore $Cl(7)$ will be isomorphic to complex split octonionic chains. Just like $e_1e_2e_3$ commutes with all elements in $Cl(3)$, $\overleftarrow{e_1e_2e_3e_4e_5e_6e_7}$ will commute with all the elements of $Cl(7)$, and it squares to 1.
We can now make particles from the left set and the right set in a similar way as is done in $Cl(6)$. The MTIS for the left set will be $\alpha_1 = \frac{-e_5 + ie_4}{2}, \alpha_2 = \frac{-e_3 + ie_1}{2}, \alpha_3 = \frac{-e_6 + ie_2}{2}$. The idempotent will be $\Omega_L\Omega_L^{\dagger} = \alpha_1\alpha_2\alpha_3\alpha_3^{\dagger}\alpha_2^{\dagger}\alpha_1^{\dagger}$. The left-handed neutrino family and their right-handed anti-particles are:
\begin{align}
    \overline{\mathcal{V}} &= \frac{ie_8 + 1}{2} & \mathcal{V} &= \frac{-ie_8 + 1}{2}\\
    V_{ad1} &= \frac{(e_5 + ie_4)}{2} & V_{d1} &= \frac{(e_5 - ie_4)}{2}\\
    V_{ad2} &= \frac{(e_3 + ie_1)}{2} & V_{d2} &= \frac{(e_3 - ie_1)}{2}\\
    V_{ad3} &= \frac{(e_6 + ie_2)}{2} & V_{d3} &= \frac{(e_6 - ie_2)}{2}\\
    V_{u1} &= \frac{(e_4 + ie_5)}{2} & V_{au1} &= \frac{(e_4 - ie_5)}{2}\\
    V_{u2} &= \frac{(e_1 + ie_3)}{2} & V_{au2} &= \frac{(e_1 - ie_3)}{2}\\
    V_{u3} &= \frac{(e_2 + ie_6)}{2} & V_{au3} &= \frac{(e_2 - ie_6)}{2}\\
    V_{e+} &= -\frac{(i + e_8)}{2} & V_{e-} &= -\frac{(-i + e_8)}{2}
\end{align}
The MTIS for the right set will be $\alpha_1 = -\omega\frac{-e_5 + ie_4}{2}$, $\alpha_2 = -\omega\frac{-e_3 + ie_1}{2}$, $\alpha_3 = -\omega\frac{-e_6 + ie_2}{2}$. The idempotent will be $\Omega_R\Omega_R^{\dagger} = \alpha_1\alpha_2\alpha_3\alpha_3^{\dagger}\alpha_2^{\dagger}\alpha_1^{\dagger}$. The right-handed neutrino family and their left-handed anti particles are:
\begin{align}
    \overline{\mathcal{V}} &= \frac{ie_8 + 1}{2} & \mathcal{V} &= \frac{ie_8 + 1}{2}\\
    V_{ad1} &= \omega\frac{(-e_5 - ie_4)}{2} & V_{d1} &= \omega\frac{(-e_5 + ie_4)}{2}\\
    V_{ad2} &= \omega\frac{(-e_3 - ie_1)}{2} & V_{d2} &= \omega\frac{(-e_3 + ie_1)}{2}\\
    V_{ad3} &= \omega\frac{(-e_6 - ie_2)}{2} & V_{d3} &= \omega\frac{(-e_6 + ie_2)}{2}\\
    V_{u1} &= \frac{(e_4 + ie_5)}{2} & V_{au1} &= \frac{(e_4 - ie_5)}{2}\\
    V_{u2} &= \frac{(e_1 + ie_3)}{2} & V_{au2} &= \frac{(e_1 - ie_3)}{2}\\
    V_{u3} &= \frac{(e_2 + ie_6)}{2} & V_{au3} &= \frac{(e_2 - ie_6)}{2}\\
    V_{e+} &= \omega\frac{(i + e_8)}{2} & V_{e-} &= \omega\frac{(-i + e_8)}{2}
\end{align}
Therefore from the two sets on $Cl(7)$ we have one generation of left-right symmetric fermions. Both the right sector and the left sector have $SU(3) \times U(1)$ symmetry.

We have discussed earlier that the right action of $\Omega$ and $\Omega^{\dagger}$ reverses the isospin of particles. The right action of $\mathbb{C}\otimes \mathbb{H}$ on $Cl(3)$ and $Cl(7)$ changes the chirality and isospin of the particles \cite{f1, f2}. In doing so we map to the Clifford algebras $Cl(5)$ and $Cl(9)$ respectively.
The right action of $\mathbb{C}\otimes\mathbb{H}$ on $\mathbb{C}\otimes\mathbb{H}$ will look like
\begin{equation}
    (1, e_1, e_2, e_1e_2)\otimes_C(1, e_1, e_2, e_1e_2)
\end{equation}
This will give us the $Cl(4)$ algebra, which we can generate using the vectors
\begin{equation}
    \{\tau_1ie_2, \tau_2ie_2, \tau_3ie_2, ie_1\}
\end{equation}
Here $\tau_1 = \Omega_L + \Omega_L^{\dagger}$, $\tau_2 = i\Omega_L - i\Omega_L^{\dagger}$, $\tau_3 = \Omega_L\Omega_L^{\dagger} - \Omega_L^{\dagger}\Omega_L$. We note that the MTIS in this case will be two dimensional spanned by
\begin{equation}
    \beta_1 = \frac{-e_1 + ie_2\tau_3}{2}, \quad \beta_2 = \Omega_L^{\dagger}ie_2
\end{equation}
The idempotent will be $\beta_1^{\dagger}\beta_2^{\dagger}\beta_2\beta_1 = \mathcal{V}_R$. We can obtain the right ideals by right multiplication of $Cl(4)$ on this ideal. We identify the following particles
\begin{align}
    \beta_1^{\dagger}\beta_2^{\dagger}\beta_2\beta_1 = \mathcal{V}_R; \quad e^{-}_R = \mathcal{V}_R\beta_1^{\dagger}\beta_2^{\dagger}\\
    e^{-}_L = \mathcal{V}_R\beta_2^{\dagger}; \quad \mathcal{V}_L = \mathcal{V}_R\beta_1^{\dagger}
\end{align}
The SU(2) symmetry can be generated by the following three generators
\begin{equation}
    T_1 = \tau_1\frac{1 + ie_1e_2}{2}, T_2 = \tau_2\frac{1 + ie_1e_2}{2}, T_3 = \tau_3\frac{1 + ie_1e_2}{2}
\end{equation}
It is interesting to note that the SU(2) operators will annihilate the particles $\mathcal{V}_R$ and $e_R^-$, whereas it will interchange
$e_L^-$ and $\mathcal{V}_L$ as isospin states. Therefore, $SU(2)_L$ acts on left-handed particles or conversely on right-handed anti-particles.
Similarly we can understand the right action of $\mathbb{C}\otimes\mathbb{H}$ on $\omega \mathbb{C}\otimes\mathbb{H}$. This can be written as
\begin{equation}
    (1, -e_1, -e_2, -e_1e_2)\otimes_C(1, e_1, e_2, e_1e_2)
\end{equation}
The algebra will again be $Cl(4)$, the particles generated through similar analysis will be
\begin{align}
    \beta_1^{\dagger}\beta_2^{\dagger}\beta_2\beta_1 = \mathcal{V}_L; \quad e^{-}_L = \mathcal{V}_L\beta_1^{\dagger}\beta_2^{\dagger}\\
    e^{-}_R = \mathcal{V}_L\beta_2^{\dagger}; \quad \mathcal{V}_R = \mathcal{V}_L\beta_1^{\dagger}
\end{align}
The $SU(2)$ generators will now annihilate $\mathcal{V}_L$ and $e_L^-$ (eq. 54), whereas it will interchange
$e_R^-$ and $\mathcal{V}_R$ as isospin states. Therefore, $SU(2)_R$ acts on right-handed particles or conversely on left-handed anti-particles. It is once again important to emphasise that the MTIS for $SU(2)$, both for the left-handed and the right-handed case, is being built from the generators of $SU(3)$.

\section{Three fermion generations, and interpretation of the right-handed sector as pre-gravitation}

The introduction of the right-handed sterile neutrinos as a part of the right-handed sector is an unambiguous prediction of our work, and  provides an important conceptual hint towards gravitation. Sterile neutrinos interact only via gravitation, and moreover only quantum mechanically. Therefore, our theory having sterile neutrinos, if complete and consistent, must also incorporate quantum gravity, and that too in the sense that it should be a formulation of quantum theory without classical time. And since mass is a source of gravitation, the theory should also explain the observed mass ratios of quarks and charged leptons over all three fermion generations, which it does \cite{vvs}.
To proceed further, we also ask why  the square-root mass ratio 3: 2: 1 of the down quark, up quark and electron is in the reverse order of their electric charge ratio  1: 2: 3? We believe that rather than being a coincidence, this fact points to deep physics, and that the symmetry group $E_6$ has an answer. We will assume that physical space is an eight dimensional space labeled by the octonions. Three generations of fermions reside on this space-time on which $E_6$ acting as the symmetry group is a candidate for the unification of the standard model with gravity, as we now argue.  The left-handed fermions are argued to be charge eigenstates, whereas the right-handed fermions are square-root mass eigenstates. The eigenvalues of the exceptional Jordan algebra, along with the corresponding eigen-matrices, permit the expression of charge eigenstates in terms of square-root mass eigenstates, and hence can be used to deduce mass ratios of charged fermions \cite{vvs}.

$E_6$ is the only exceptional Lie group which has complex representations, and it has two  maximal subgroups $\tilde{H_1} = [SU(3)\times SU(3)\times SU(3)]/\mathbf{Z_3}$, $\tilde{H_2} = Spin(10)$. Their intersection is $SU(3)\times SU(2)_R\times SU(2)_L\times U(1)$ which is the gauge-group for left-right symmetric Pati-Salam model. The groups belonging to the two maximal sub-groups but lying outside the intersection are $Spin(6)$ and $SU(3)\times SU(3)$. We identify one of these two $SU(3)$ with generational symmetry, giving rise to the three fermion generations,  and now the novel part is that we introduce gravi-color, analogous to QCD color, and associate this third $SU(3)$ with pre-gravitation and square-root mass number, in lieu of electric charge. This will help understand the observed down quark : up quark : electron square-root mass ratio of 3: 2: 1 Just as $SU(3)_c\times U(1)_{em}$ is described by the Clifford algebra $Cl(6)$ as unbroken electro-color, the group $SU(3)_{grav}\times U(1)_g$ will describe unbroken gravi-color through another copy of $Cl(6)$ and together these two copies of $Cl(6)$ will form a $Cl(7)$ using the complex split bioctonions as analysed above. This offers a unification of QCD color with gravi-color, prior to the L-R symmetry breaking, which we assume is the same as the electro-weak symmetry breaking. The group $SU(2)_L \times SU(2)_R$ describes gravi-weak unification through complex split biquaternions; $SU(2)_L$ is the standard model weak symmetry and $SU(2)_R$ is the gravi- part of gravi-weak, mediated by two  spin one gravitationally charged `Lorentz' bosons, and the Higgs. In our theory there are no fundamental gravitons; these are replaced by two right-handed Lorentz bosons. From here, gravitation as a second-rank tensor field theory emerges in the classical approximation.  The electro-weak symmetry breaking also breaks the gravi-weak symmetry. The $Spin(6)$ which is not in the intersection is identified as a six dimensional Minkowski spacetime because of the isomorphism $Spin(6)\sim SO(5,1)\sim SL(2,H)$. This possibly is the space-time spanned by the gravi-weak interaction. Our classical space-time is 4D Minkowski and the fifth and sixth dimension are of the order of thickness of the range of the weak interaction: the weak symmetry is a space-time symmetry as well as an internal symmetry.

Prior to L-R symmetry breaking, the neutrino is a Dirac neutrino, which after symmetry breaking separates into the left-handed active Majorana neutrino, and the right-handed sterile Majorana neutrino. Analogous to how it was done above, we use the Dirac neutrino as an idempotent, prior to L-R symmetry breaking, and construct the Clifford algebra $Cl(7)=Cl(6) \oplus Cl(6)$ displayed below.
\begin{align}
    {\mathcal{\overline V}_L} &= \frac{ie_8 + 1}{2} & \mathcal{\overline V}_R &= \frac{ie_8 + 1}{2}\\
    V_{ad1} &= \frac{(e_5 + ie_4)}{2} & V_{e+1} &= \omega\frac{(-e_5 - ie_4)}{2}\\
    V_{ad2} &= \frac{(e_3 + ie_1)}{2} & V_{e+2} &= \omega\frac{(-e_3 - ie_1)}{2}\\
    V_{ad3} &= \frac{(e_6 + ie_2)}{2} & V_{e+3} &= \omega\frac{(-e_6 - ie_2)}{2}\\
    V_{u1} &= \frac{(e_4 + ie_5)}{2} & V_{au1} &= \frac{(e_4 + ie_5)}{2}\\
    V_{u2} &= \frac{(e_1 + ie_3)}{2} & V_{au2} &= \frac{(e_1 + ie_3)}{2}\\
    V_{u3} &= \frac{(e_2 + ie_6)}{2} & V_{au3} &= \frac{(e_2 +ie_6)}{2}\\
    V_{e+} &= -\frac{(i + e_8)}{2} & V_{ad} &= \omega\frac{(i + e_8)}{2}
\end{align}
The eight fermions on the left are made by using the left-handed anti-neutrino as the idempotent, while the eight fermions on the right are made by using the right-handed anti-neutrino as idempotent. The two sets share a common number  $U(1)_{electro-gravi}$ operator defined as usual by
\begin{equation}
     Q_{gem} = \frac{\alpha_1^{\dagger}\alpha_1 + \alpha_2^{\dagger}\alpha_2 + \alpha_3^{\dagger}\alpha_3}{3}
\end{equation}
and have an $SU(3)_c \times SU(3)_{grav}$ symmetry, which we interpret as the unification of QCD color and `would-be-gravity', and also of electromagnetism and a $U(1)_{grav}$.  Here, $ Q_{gem}$  is the gravi-electric.charge number operator: after the symmetry breaking this will be interpreted as the electric charge for the left-handed particles, and square-root mass number for the right handed particles. The $U(1)_{electro-gravi}$ boson will separate into the photon for electromagnetism, and a newly proposed gravitational boson. Prior to symmetry breaking the particle content for one generation is as follows. Anti-particles are obtained by ordinary complex conjugation of the particles, as before.
The Dirac neutrino is the sum of the left handed neutrino and the right handed neutrino; it has $Q_{gem}=0$, is a singlet under $SU(3)_c \times SU(3)_{grav}$ and we can denote it as the particle LeftHandedNeutrino-RightHandedNeutrino, and after the L-R symmetry breaking it acquires mass and separates into a left-handed active Majorana neutrino and a right handed sterile Majorana neutrino.
The first excitation above the idempotent has $Q_{gem}=1/3$ and is an anti-triplet under $SU(3)_c$ and an anti-triplet under $SU(3)_{grav}$. We denote this particle as LeftHandedAntiDownQuark-RightHandedPositron. After the L-R symmetry breaking it separates into the left-handed anti-down quark of electric charge $1/3$ and right-handed positron of square-root mass number $1/3$ 
The second excitation above the idempotent has $Q_{gem}=2/3$ and is a triplet under $SU(3)_c$ and  a triplet under $SU(3)_{grav}$. We denote this particle as LeftHandedUpQuark-RightHandedUpQuark. After the L-R symmetry breaking it separates into the left-handed up quark of electric charge $2/3$ and right-handed up quark of square-root mass number $2/3$. 
The third excitation above the idempotent has $Q_{gem}=1$ and is a singlet under both $SU(3)_c$ and $SU(3)_{grav}$. We denote this particle as LeftHandedPositron-RightHandedAntiDownQuark. After the L-R symmetry breaking it separates into a left-handed positron of electric charge 1 and a right-handed anti-down quark of  square-root mass number 1.
The corresponding anti-particles have a $Q_{gem}$ number of the opposite sign.

We propose to identify the right-handed positron of square-root mass number 1/3 with the left-handed positron of electric charge 1 as being the same particle. This is essentially a proposal for a gauge-gravity duality which we have justified from the dynamics \cite{Singhreview}. Similarly, the right-handed anti-down quark with square-root mass number 1 is identified with the left-handed anti-down quark of electric charge 1/3. The right-handed up quark of square-root mass number 2/3 is identified with the left-handed up quark of electric charge 2/3. In this way we recover one generation of standard model fermions after the L-R symmetry breaking. An analogous correspondence is obtained for the other two fermion generations.

Before symmetry breaking, we can define $\ln \alpha_{unif}\propto 2\ln Q_{gem}\equiv \ln(AB) = \ln A + \ln B \propto e + \sqrt{m}$ where $\ln A$ is proportional to electric charge and $\ln B$ is proportional to square-root mass number, and at the time of L-R symmetry breaking $2Q_{gem}$ separates into two equal parts, one identified with electric charge, and the other with square-root mass. We hence see that in the unified L-R phase we can define a new entity, a charge-root-mass as $\alpha_{unif} = \exp e \exp\sqrt{m}\equiv E\sqrt{M}$. This is the source of the unified force described by a $U(1)$ boson, sixteen gravi-gluons, and six gravi-weak bosons corresponding to $SU(2)_L \times SU(2)_R$ and the Higgs; adding to a total of 24 bosons. There are 48 fermions for three generations, giving a total of 48+24 =72, to which if we add six d.o.f. for the six dimensional space-time $SO(1,5)$ we might be able to account for the 78 dimensional $E_6$. The gravi-weak bosons generate the Lorentz-weak symmetry by their right action on the $Cl(7)$, as described in \cite{Vatsalya1}. After symmetry breaking this separates into the short range weak interaction and long-range gravity described by general relativity. $SU(3)_{grav}$ is negligible in strength  compared to QCD color but plays a very important role of describing the square-root mass number as source of would-be-gravity and showing that mass-quantisation arises only after the standard model has been unified with gravity, as was always anticipated. We also see via $E_6$ that $SU(3)_{grav} \times SU(2)_R \times U(1)_g$ is the pre-gravitational counterpart of the standard model $SU(3)_{c} \times SU(2)_L \times U(1)_{em}$. The remaining entities from the two maximal sub-groups, i.e. $SU(3)_{gen}$ and $Spin(6)$ respectively give rise to three generations and a 6D Minkowski space-time.
We now finally understand why the square-root mass ratios 3:2:1 for down : up : electron are in the reverse order as the ratio $1 : 2 : 3$ of their electric charge. It is a consequence of the gauge-gravity duality afforded by $E_6$. Also, classical gravitation as we know it is emergent from this above-mentioned pre-gravitation.


In summary, in the preceding sections we have argued that elementary particles live in a space labelled by the octonions and having the symmetry group $E_6$. Doing so reveals the symmetries of the standard model, the quantisation of electric charge and mass, and the relation between the standard model and pre-gravitation and space-time, through the extension to the left-right symmetric model. We now discuss how dynamics can be set up on this space, and how quantum theory and classical spacetime can be recovered.

\section{A matrix-valued Lagrangian dynamics as a pre-spacetime, pre-quantum dynamics}
 As we have noted, there ought to exist a reformulation of quantum [field] theory which does not depend on classical time. To arrive at such a reformulation, valid at the Planck time scale resolution, one constructs a  pre-quantum,  pre-spacetime theory, in two steps. In the first step, the pre-quantum theory is constructed by retaining classical space-time, and raising classical matter configuration variables and their corresponding canonical momenta to the status of matrices [equivalently, operators]. A matrix-valued Lagrangian dynamics is constructed, with the Lagrangian being the trace of a matrix polynomial. The matrices themselves have complex  Grassmann numbers as entries, with even-grade Grassmann matrices $q_B$(odd-grade matrices $q_F$) representing bosonic(fermionic) degrees of freedom. Matrix-valued equations of motion are derived from the trace Lagrangian, but the Heisenberg algebra is not imposed on the matrix commutators. Instead, the commutators evolve dynamically, and yet the global unitary invariance of the trace Hamiltonian implies  the existence of the following novel conserved charge, made from the commutators and anti-commutators:
\begin{equation}
\tilde{C} = \sum_i [q_{Bi}, p_{Bi}] - \sum_j \{q_{Fj}, p_{Fj}\}
\label{cons}
\end{equation} 
This is Adler's trace dynamics \cite{Adler:04}  assumed operational at the Planck time scale resolution, and the Hamiltonian of the theory is in general not self-adjoint. Next, we ask as to what is the emergent 
coarse-grained dynamics at low energies, if the trace dynamics is not being examined at Planck time resolution? If not too many degrees of freedom are entangled with each other, then the anti-self-adjoint part of the Hamiltonian is negligible, and the emergent dynamics is quantum [field] theory. The conserved charge mentioned above gets equipartitioned, and each commutator and anti-commutator in it is identified with $i\hbar$, where $\hbar$ is Planck's constant. On the other hand, if sufficiently many degrees of freedom are entangled, the anti-self-adjoint part of the Hamiltonian becomes significant, `collapse of the wave function' [spontaneous localisation] ensues, and ordinary classical dynamics is recovered. Qualitatively speaking, spontaneous localisation randomly maps each canonical matrix  to one or the other of its eigenvalues [Born probability rule is obeyed], and the conserved charge above goes to zero identically. Thus the underlying trace dynamics is a {\it deterministic  but non-unitary} pre-quantum theory [on a classical space-time background], and quantum field theory, along with its indeterminism,  is an emergent low-energy  thermodynamic phenomenon. The emergent indeterminism is a consequence of the coarse-graining: i.e.  not having precise information about the dynamics on  the small time-scales that have been averaged over.

In the second step, one goes from the above pre-quantum theory to a pre-quantum, pre-spacetime theory \cite{Singh2020DA}  by raising space-time points also to the status of matrices, via the following profound theorem \cite{Chams:1997}  in Riemannian geometry. The Einstein-Hilbert action is proportional to the trace of the regularised Dirac operator $D_B$ on the manifold, in a truncated heat-kernel expansion in $L_P^{-2}$: 
\begin{equation}
Tr \; [L_P^2 \; D_B^2] \sim \int d^4x \sqrt{g} \; \frac{R}{L_P^2} + {\cal O}(L_P^0)\sim L_P^2 \sum_n \lambda_n^2
\label{grraction}
\end{equation}
The eigenvalues $\lambda_n$ of the Dirac operator play the role of the dynamical variables of general relativity \cite{Rovelli}.
To go from here to the pre-quantum, pre-spacetime theory, every eigenvalue  $\lambda_n$ is raised to the status of a canonical matrix momentum: $\lambda_n\rightarrow p_{Bn}\propto q_{Bn}/d\tau \equiv D_B$, where the bosonic matrix $q_B$ is the corresponding newly introduced configuration variable.  We hence have $N$ copies of the Dirac operator ($n$ runs from 1 to $N$,  with  $N\rightarrow \infty$).  The matrix dynamics trace Lagrangian [space-time part] for the $n$-th degree of freedom is then $Tr \; (dq_{Bn}/d\tau)^2$ and the total matrix dynamics action [space-time part] is  $S \sim \sum_n \int d\tau\; Tr\; (dq_{Bn}/d\tau)^2$. Here, $\tau$ is an absolute time-parameter, known as Connes time, which is a unique feature of a {\it non-commutative} geometry [which is what we now have]  and is a consequence of the Tomita-Takesaki theory. Yang-Mills fields are represented by the matrices $q_{Bn}$, gravitation by the $\dot{q}_{Bn}$,  fermionic degrees of freedom by  fermionic matrices $q_{Fn}$ and their `velocities' $\dot{q}_{Fn}$, where `dot' denotes a derivative with respect to the time $c\tau$. 
Each of the $n$ degrees of freedom has an action of its own, which is given  by \cite{Singh2020DA}
\begin{equation}
\frac{S}{\hbar} =  \frac{a_0}{2} \int \frac{d\tau}{\tau_{Pl}} \; Tr  \bigg[\dot{q}_B^{\dagger} + i\frac{\alpha}{L} q_B^\dagger+ a_0 \beta_1\left( \dot{q}_F^\dagger  + i\frac{\alpha}{L} q_F^\dagger\right)\bigg] \times \bigg[ \dot{q}_B + i\frac{\alpha}{L} q_B+ a_0 \beta_2\left( \dot{q}_F + i\frac{\alpha}{L} q_F\right)\bigg] 
\label{ymi}
\end{equation} 
where  $a_0 \equiv L_P^2 / L^2$. The total action of this generalised trace dynamics is the sum over $n$ of $N$ copies of the above action, one copy for each degree of freedom. This total action defines the pre-spacetime, pre-quantum theory, with each degree of freedom [defined by the above action] being an `atom' of space-time-matter [an STM atom, or an `aikyon']. In an aikyon, one loses the distinction between space-time and matter: the fermionic part $q_F$ (say an electron) is the source for the bosonic part $q_B$; however the interpretation can also be reversed: $q_B$ can be thought of as the source for $q_F$. In this action, there are only three fundamental constants: Planck length, Planck time, and Planck's constant $\hbar$. $L$ is a length parameter [scaled with respect to $L_P$; $q_B$ and $q_F$ have dimensions of length] characterising the STM atom, and $\alpha$ is the dimensionless Yang-Mills coupling constant. $\beta_1$ and $\beta_2$ are two unequal complex Grassmann numbers: their being equal gives inconsistent equations of motion. The aikyon is therefore a 2D object.

The further analysis of this pre-space-time, pre-quantum theory proceeds just as for the pre-quantum trace dynamics. Equations of motion are derived, and there is a conserved charge just as in Eqn. (\ref{cons}). Assuming that the theory holds at the Planck scale, the emergent low-energy approximation obeys quantum commutation rules, and  is the sought for reformulation of quantum theory without classical time. This emergent theory is also a quantum theory of gravity. If sufficiently many aikyons get entangled, the anti-self-adjoint part of the Hamiltonian becomes significant, spontaneous localisation results, and the fermionic part of the entangled STM atoms is localised. There emerges a 4D classical space-time manifold (labelled by the positions of collapsed fermions), sourced by point masses and gauge fields, and  whose geometry obeys the laws of general relativity; [space-time from collapse of the wave-function]. Those aikyons which are not sufficiently entangled remain quantum; their dynamics is described by quantum field theory on  space-time background generated by the entangled, collapsed fermions [the macroscopic bodies of the universe]. 

The action (\ref{ymi}) reveals a great deal of new information, as we now unravel. To begin with, since we have a fundamental  Lagrangian dynamics, we should be able to define spin [angular momentum canonical to some appropriate angle] for the bosonic and fermionic degrees of freedom, and prove the spin-statistics theorem. After defining new dynamical variables $\dot{\widetilde Q}_B$ and $\dot{\widetilde{Q}}_F$ as
\begin{equation}
{\dot{\widetilde{Q}}_B} \equiv \frac{1}{L} (i\alpha q_B + L \dot{q}_B); \qquad  {\dot{\widetilde{Q}}_F} \equiv \frac{1}{L} (i\alpha q_F + L \dot{q}_F)
\label{lilqu}
\end{equation}
the Lagrangian in (\ref{ymi}) can be brought to the elegant and revealing form, akin to  a free particle:
\begin{equation}
 \mathcal{L} = \frac{a_0}{2} \; Tr  \biggl(\biggr. \dot{\widetilde{Q}}_{B}^\dagger + \dfrac{L_{p}^{2}}{L^{2}} \beta_{1} \dot{\widetilde{Q}}_{F}^{\dagger} \biggl.\biggr) \biggl(\biggr. \dot{\widetilde{Q}}_{B} + \dfrac{L_{p}^{2}}{L^{2}} \beta_{2} \dot{\widetilde{Q}}_{F} \biggl.\biggr) \biggl.
\label{eq:tracelagn}
\end{equation}
Next we introduce self-adjoint bosonic operators $R_B$ and $\theta_B$, and self-adjoint fermionic operators $R_F$ and $\theta_F$, as follows:
$\widetilde{Q}_B \equiv R_B\; \exp i\theta_B \ ;  \  \widetilde{Q}_F \equiv R_F \; \exp i\eta \theta_F$.
Here, $\eta$ is a real Grassmann number, introduced to ensure that the fermionic phase is bosonic, so that $\widetilde{Q}_F$ comes out fermionic, as desired, upon the Taylor expansion of its phase. The canonical angular momenta corresponding to the angles $\theta_B$ and $\theta_F$ are bosonic and fermionic spin, which obeys the spin-statistics theorem, and agrees with our conventional understanding of spin in relativistic quantum mechanics \cite{Singhspin1}.

But in which space  are these  angles $\theta_B$ and $\theta_F$ located? They cannot be in 4-D space-time because then the canonical momentum will be orbital angular momentum, not spin. We note from Eqn. (\ref{lilqu}) that the velocity $\dot{q}_B$ and the configuration variable $q_B$ together define a complex plane: spin measures the angular momentum of the periodic motion of $\widetilde{Q}_B$ in this plane. Making the plausible assumption that the velocity $\dot{q}_B$ has four components, which lie in our 4D space-time, we are compelled to conclude that the $q_B$ also has four components, which lie on four `internal' directions, thus making physical space eight dimensional! We already know that the $q_B$ describe Yang-Mills fields: so we have at hand a Kaluza-Klein type of theory, and spin describes periodic motion from 4D space-time to and from the internal directions. Furthermore, the coordinates in this 8D physical space will be assumed to be non-commuting,  consistent with the fact that the matrices $q_B$ and $\dot{q}_B$ do not commute. The introduction of an 8D non-commuting coordinate system immediately suggests the  octonions. The automorphisms of the octonions take over the role of space-time diffeomorphisms and internal gauge transformations, thus unifying the latter two. The [non-associative] algebra of the octonion automorphisms forms the smallest of the five exceptional Lie groups $G_2$, which has fourteen generators. The unitary transformations generated by these act on the $Q$-matrices while leaving the trace Lagrangian invariant, and once again yield the conserved charge of Eqn. (\ref{cons}). An aikyon described by the Lagrangian (\ref{ymi}) evolves in this eight dimensional octonionic space in Connes time $\tau$.  Spontaneous localisation confines classical systems to a 4D Minkowski space-time obeying general relativity, whereas quantum systems reside in the full 8D space,  though they can be described from a 4D space-time perspective as well, provided we adopt the conventional definition of quantum spin. We have already seen that, even before dynamics is invoked, the octonion space already dictates the symmetries of the standard model, including Lorentz symmetry, and the quantisation of electric charge and of fermion masses. When the $q$ matrix components are much smaller than unity, in Planck units, we have an IR situation: however, and this is important, the fundamental description must still be on an octonionic space. The relation of this Lagrangian with the algebraic description of the standard model has been discussed in detail in \cite{Singhreview}, including how the Lagrangian together with the algebra of the octonions leads to a theoretical derivation of the low energy fine structure constant.

The matrices $\widetilde{Q}_B$ and $\widetilde{Q}_F$ both have eight components each, one component per octonionic direction. We can view this as a generalisation of Newtonian dynamics, where the 3D space  labeled by real numbers has been replaced by an 8D space labeled by the octonions. The real number valued configuration variables have been replaced by matrices. Absolute time is still there as Connes time, but the Hamiltonian is no longer self-adjoint. Overall, if we think of $L_P/L$ as representing square-root of mass, this action principle has the same form as that for a free particle in Newtonian dynamics. It is gratifying to consider the possibility that this transition to a more general [deterministic but non-unitary]  Newtonian dynamics could capture known physical phenomena of quantum nature, as well as gravitation. Interaction between different STM atoms is through collisions in the octonion space and through entanglement.

\section{Interpretation}
The dynamics of a quantum system in flight, say an electron, should be describable without making any reference to classical time. We propose that such a description is given by the action (\ref{ymi}) and by the Lagrangian dynamics implied by this action, with evolution taking place in Connes time. We do not not think of this evolution as taking place in a spacetime, but rather in an octonionic space having the symmetry $E_6$. This is true at all energy scales, including at the low energies at which we currently observe the elementary particles described by the standard model. Moreover, the system possesses left-right symmetry during flight. The algebra of the octonionic space determines the  symmetries of the standard model, and the values of the parameters of the standard model and mass ratios of quarks and charged leptons. These parameters are not being fixed by some yet unknown high energy physics, but by the quantum structure of spacetime, which is non-trivial even at low energies. A quantum system carries its `quantum spacetime' with it, and that is what determines the low energy standard model. At these low energies the $q$-matrices which describe the dynamics are close to identity, and do not significantly distort the octonionic space, which is the non-commutative counterpart of Minkowski spacetime. But the octonion space has structure, by virtue of its non-commutativity and non-associativity, and this structure is very restrictive. One can no longer assign arbitrary properties to fermions on such a space.
\begin{figure}[h]
\centering
\includegraphics[width=12cm]{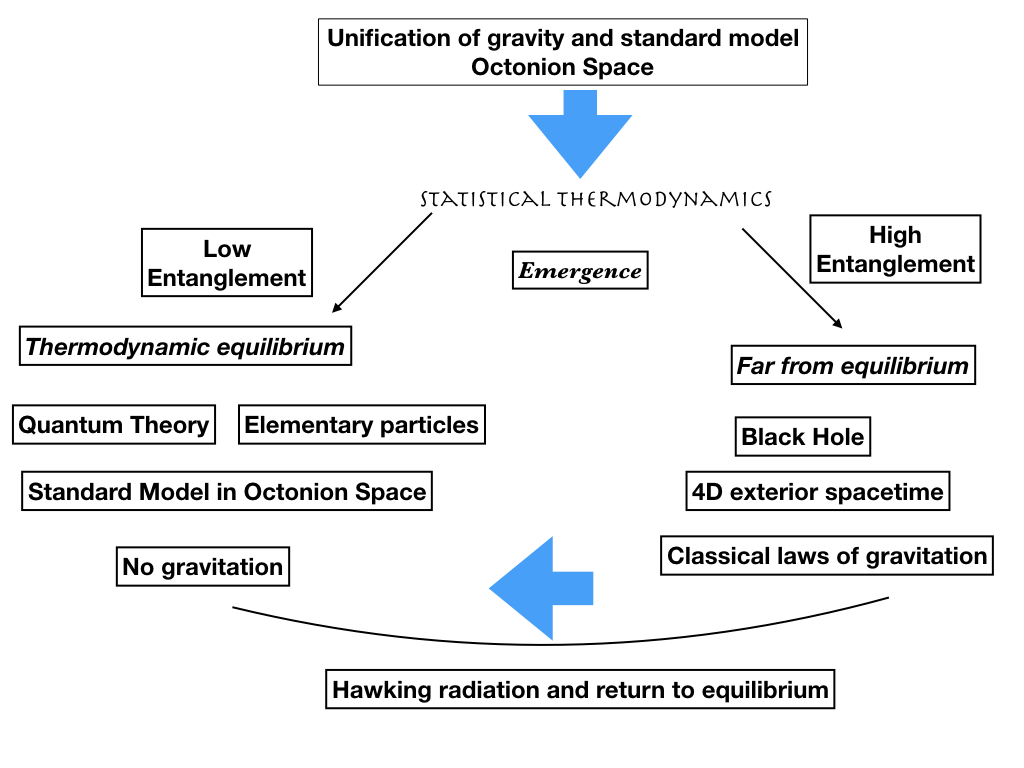}
\caption{Quantum theory, and gravitation, as emergent phenomena. In this diagram there is no reference to an energy scale, except to note that the fundamental theory holds at some time resolution scale which has not yet been probed by experiments. Coarse-graining over that time scale yields quantum theory on an octonionic space, in the low entanglement limit. The parameters and the symmetries of the standard model are no longer arbitrary. In the high entanglement limit classical space-time and classical gravitation emerge.}
\end{figure}
Classical macroscopic bodies arise when sufficiently many quantum degrees of freedom get entangled - as a result of such super-critical entanglement, the anti-self-adjoint part of the Hamiltonian becomes significant, resulting in spontaneous localisation, and the emergence of classical space-time geometry. Quantum to classical transition also breaks the left-right symmetry. On the classical background the dynamics given by (\ref{ymi}) becomes equivalent to standard quantum field theory to a great accuracy. However when we work directly with quantum field theory on a Minkowski background, which only arises after coarse-graining, we miss out on  the crucial information the more precise underlying theory is trying to give us. It is the same as not being able to understand the macroscopic properties of a fluid such as water, which is what happens if we ignore the microscopic properties of water at the molecular level. And molecular behaviour determines macroscopic properties even though the two associated length scales are extremely different. In much the same way, quantum theory and gravitation both are emergent phenomena. We illustrate this in the diagram shown  in Fig. 2.

Further work is in progress to put these claims on a firmer footing.

\section{Author Declarations}

\noindent {\bf Conflict of Interest: } The author has no conflicts of interest to disclose.

\noindent{\bf Data Availability:} The data that supports the findings of this article are available within the article.


\bibliography{biblioqmtstorsionn}

\def\polhk#1{\setbox0=\hbox{#1}{\ooalign{\hidewidth
  \lower1.5ex\hbox{`}\hidewidth\crcr\unhbox0}}} \def\cprime{$'$}
  \def\cprime{$'$}
\begin{thebibliography}{36}%
\makeatletter
\providecommand \@ifxundefined [1]{%
 \@ifx{#1\undefined}
}%
\providecommand \@ifnum [1]{%
 \ifnum #1\expandafter \@firstoftwo
 \else \expandafter \@secondoftwo
 \fi
}%
\providecommand \@ifx [1]{%
 \ifx #1\expandafter \@firstoftwo
 \else \expandafter \@secondoftwo
 \fi
}%
\providecommand \natexlab [1]{#1}%
\providecommand \enquote  [1]{``#1''}%
\providecommand \bibnamefont  [1]{#1}%
\providecommand \bibfnamefont [1]{#1}%
\providecommand \citenamefont [1]{#1}%
\providecommand \href@noop [0]{\@secondoftwo}%
\providecommand \href [0]{\begingroup \@sanitize@url \@href}%
\providecommand \@href[1]{\@@startlink{#1}\@@href}%
\providecommand \@@href[1]{\endgroup#1\@@endlink}%
\providecommand \@sanitize@url [0]{\catcode `\\12\catcode `\$12\catcode
  `\&12\catcode `\#12\catcode `\^12\catcode `\_12\catcode `\%12\relax}%
\providecommand \@@startlink[1]{}%
\providecommand \@@endlink[0]{}%
\providecommand \url  [0]{\begingroup\@sanitize@url \@url }%
\providecommand \@url [1]{\endgroup\@href {#1}{\urlprefix }}%
\providecommand \urlprefix  [0]{URL }%
\providecommand \Eprint [0]{\href }%
\providecommand \doibase [0]{http://dx.doi.org/}%
\providecommand \selectlanguage [0]{\@gobble}%
\providecommand \bibinfo  [0]{\@secondoftwo}%
\providecommand \bibfield  [0]{\@secondoftwo}%
\providecommand \translation [1]{[#1]}%
\providecommand \BibitemOpen [0]{}%
\providecommand \bibitemStop [0]{}%
\providecommand \bibitemNoStop [0]{.\EOS\space}%
\providecommand \EOS [0]{\spacefactor3000\relax}%
\providecommand \BibitemShut  [1]{\csname bibitem#1\endcsname}%
\let\auto@bib@innerbib\@empty
\bibitem [{\citenamefont {Penrose}(1996)}]{Penrose:96}%
  \BibitemOpen
  \bibfield  {author} {\bibinfo {author} {\bibfnamefont {Roger}\ \bibnamefont
  {Penrose}},\ }\bibfield  {title} {\enquote {\bibinfo {title} {On gravity's
  role in quantum state reduction},}\ }\href@noop {} {\bibfield  {journal}
  {\bibinfo  {journal} {Gen. Rel. Grav.}\ }\textbf {\bibinfo {volume} {28}},\
  \bibinfo {pages} {581--600} (\bibinfo {year} {1996})}\BibitemShut {NoStop}%
\bibitem [{\citenamefont {Singh}(2021{\natexlab{a}})}]{GRFEssay2021}%
  \BibitemOpen
  \bibfield  {author} {\bibinfo {author} {\bibfnamefont {Tejinder~P.}\
  \bibnamefont {Singh}},\ }\bibfield  {title} {\enquote {\bibinfo {title}
  {Quantum theory without classical time: octonions, and a theoretical
  derivation of the fine structure constant 1/137},}\ }\href@noop {} {\bibfield
   {journal} {\bibinfo  {journal} {Int. J. Mod. Phys. D}\ }\textbf {\bibinfo
  {volume} {30}},\ \bibinfo {pages} {2142010
  https://doi.org/10.1142/S0218271821420104 arXiv:2110:07548} (\bibinfo {year}
  {2021}{\natexlab{a}})}\BibitemShut {NoStop}%
\bibitem [{\citenamefont {Singh}(2021{\natexlab{b}})}]{Singhreview}%
  \BibitemOpen
  \bibfield  {author} {\bibinfo {author} {\bibfnamefont {Tejinder~P.}\
  \bibnamefont {Singh}},\ }\bibfield  {title} {\enquote {\bibinfo {title}
  {Quantum theory without classical time: a route to quantum gravity and
  unification},}\ }\href@noop {} {\bibfield  {journal} {\bibinfo  {journal}
  {arXiv:2110.02062v1}\ } (\bibinfo {year} {2021}{\natexlab{b}})}\BibitemShut
  {NoStop}%
\bibitem [{\citenamefont {Ramond}(1976)}]{Ramond1976}%
  \BibitemOpen
  \bibfield  {author} {\bibinfo {author} {\bibfnamefont {Pierre}\ \bibnamefont
  {Ramond}},\ }\bibfield  {title} {\enquote {\bibinfo {title} {Introduction to
  exceptional {L}ie groups and algebras},}\ }\href@noop {} {\bibfield
  {journal} {\bibinfo  {journal} {https://inspirehep.net/literature/111550}\
  }\textbf {\bibinfo {volume} {CALT-68-577}} (\bibinfo {year}
  {1976})}\BibitemShut {NoStop}%
\bibitem [{\citenamefont {Dixon}(1994)}]{Dixon}%
  \BibitemOpen
  \bibfield  {author} {\bibinfo {author} {\bibfnamefont {Geoffrey~M.}\
  \bibnamefont {Dixon}},\ }\href@noop {} {\emph {\bibinfo {title} {Division
  algebras, octonions, quaternions, complex numbers and the algebraic design of
  physics}}}\ (\bibinfo  {publisher} {Kluwer, Dordrecht},\ \bibinfo {year}
  {1994})\BibitemShut {NoStop}%
\bibitem [{\citenamefont {Tze}\ and\ \citenamefont {Gursey}(1996)}]{Gursey}%
  \BibitemOpen
  \bibfield  {author} {\bibinfo {author} {\bibfnamefont {C.~H.}\ \bibnamefont
  {Tze}}\ and\ \bibinfo {author} {\bibfnamefont {F.}~\bibnamefont {Gursey}},\
  }\href@noop {} {\emph {\bibinfo {title} {On the role of division, {Jordan}
  and related algebras in particle physics}}}\ (\bibinfo  {publisher} {World
  Scientific Publishing},\ \bibinfo {year} {1996})\BibitemShut {NoStop}%
\bibitem [{\citenamefont {Furey}(2015)}]{f1}%
  \BibitemOpen
  \bibfield  {author} {\bibinfo {author} {\bibfnamefont {Cohl}\ \bibnamefont
  {Furey}},\ }\bibfield  {title} {\enquote {\bibinfo {title} {Standard model
  physics from an algebra? {Ph. D.} thesis, university of {Waterloo}},}\
  }\href@noop {} {\ \textbf {\bibinfo {volume} {arXiv:1611.09182 [hep-th]}}
  (\bibinfo {year} {2015})}\BibitemShut {NoStop}%
\bibitem [{\citenamefont {Furey}(2018{\natexlab{a}})}]{f2}%
  \BibitemOpen
  \bibfield  {author} {\bibinfo {author} {\bibfnamefont {Cohl}\ \bibnamefont
  {Furey}},\ }\bibfield  {title} {\enquote {\bibinfo {title} {Three
  generations, two unbroken gauge symmetries, and one eight-dimensional
  algebra},}\ }\href@noop {} {\bibfield  {journal} {\bibinfo  {journal} {Phys.
  Lett. B}\ }\textbf {\bibinfo {volume} {785}},\ \bibinfo {pages} {1984}
  (\bibinfo {year} {2018}{\natexlab{a}})}\BibitemShut {NoStop}%
\bibitem [{\citenamefont {Furey}(2018{\natexlab{b}})}]{f3}%
  \BibitemOpen
  \bibfield  {author} {\bibinfo {author} {\bibfnamefont {Cohl}\ \bibnamefont
  {Furey}},\ }\bibfield  {title} {\enquote {\bibinfo {title} {${SU(3)_C\times
  SU(2)_L \times U(1)_Y (\times U(1)_X)}$ as a symmetry of division algebraic
  ladder operators},}\ }\href@noop {} {\bibfield  {journal} {\bibinfo
  {journal} {Euro. Phys. J. C}\ }\textbf {\bibinfo {volume} {78}},\ \bibinfo
  {pages} {375 arXiv:1806.00612 [hep--th]} (\bibinfo {year}
  {2018}{\natexlab{b}})}\BibitemShut {NoStop}%
\bibitem [{\citenamefont {Chisholm}\ and\ \citenamefont {Farwell}(1996 Ed. W.
  R. Baylis)}]{Chisholm}%
  \BibitemOpen
  \bibfield  {author} {\bibinfo {author} {\bibfnamefont {J.}~\bibnamefont
  {Chisholm}}\ and\ \bibinfo {author} {\bibfnamefont {R.}~\bibnamefont
  {Farwell}},\ }\bibfield  {title} {\enquote {\bibinfo {title} {Clifford
  geometric algebras: with applications to physics, mathematics and
  engineering},}\ \ }(\bibinfo  {publisher} {Birkhauser, Boston},\ \bibinfo
  {year} {1996 Ed. W. R. Baylis})\ p.\ \bibinfo {pages} {365}\BibitemShut
  {NoStop}%
\bibitem [{\citenamefont {Trayling}\ and\ \citenamefont
  {Baylis}(2001)}]{Trayling}%
  \BibitemOpen
  \bibfield  {author} {\bibinfo {author} {\bibfnamefont {G.}~\bibnamefont
  {Trayling}}\ and\ \bibinfo {author} {\bibfnamefont {W.}~\bibnamefont
  {Baylis}},\ }\bibfield  {title} {\enquote {\bibinfo {title} {A geometric
  basis for the standard-model gauge group},}\ }\href@noop {} {\bibfield
  {journal} {\bibinfo  {journal} {J. Phys. A: Math. Theor.}\ }\textbf {\bibinfo
  {volume} {34}},\ \bibinfo {pages} {3309} (\bibinfo {year}
  {2001})}\BibitemShut {NoStop}%
\bibitem [{\citenamefont {Dubois-Violette}(2016)}]{Dubois_Violette_2016}%
  \BibitemOpen
  \bibfield  {author} {\bibinfo {author} {\bibfnamefont {Michel}\ \bibnamefont
  {Dubois-Violette}},\ }\bibfield  {title} {\enquote {\bibinfo {title}
  {Exceptional quantum geometry and particle physics},}\ }\href {\doibase
  10.1016/j.nuclphysb.2016.04.018} {\bibfield  {journal} {\bibinfo  {journal}
  {Nuclear Physics B}\ }\textbf {\bibinfo {volume} {912}},\ \bibinfo {pages}
  {426--449} (\bibinfo {year} {2016})}\BibitemShut {NoStop}%
\bibitem [{\citenamefont {Todorov}(2019)}]{Todorov:2019hlc}%
  \BibitemOpen
  \bibfield  {author} {\bibinfo {author} {\bibfnamefont {Ivan}\ \bibnamefont
  {Todorov}},\ }\bibfield  {title} {\enquote {\bibinfo {title} {{Exceptional
  quantum algebra for the standard model of particle physics}},}\ }\href@noop
  {} {\bibfield  {journal} {\bibinfo  {journal} {Nucl. Phys. B}\ }\textbf
  {\bibinfo {volume} {938}},\ \bibinfo {pages} {751 arXiv:1808.08110 [hep--th]}
  (\bibinfo {year} {2019})}\BibitemShut {NoStop}%
\bibitem [{\citenamefont {Dubois-Violette}\ and\ \citenamefont
  {Todorov}(2019)}]{Dubois-Violette:2018wgs}%
  \BibitemOpen
  \bibfield  {author} {\bibinfo {author} {\bibfnamefont {Michel}\ \bibnamefont
  {Dubois-Violette}}\ and\ \bibinfo {author} {\bibfnamefont {Ivan}\
  \bibnamefont {Todorov}},\ }\bibfield  {title} {\enquote {\bibinfo {title}
  {{Exceptional quantum geometry and particle physics II}},}\ }\href {\doibase
  10.1016/j.nuclphysb.2018.12.012} {\bibfield  {journal} {\bibinfo  {journal}
  {Nucl. Phys. B}\ }\textbf {\bibinfo {volume} {938}},\ \bibinfo {pages}
  {751--761 arXiv:1808.08110 [hep--th]} (\bibinfo {year} {2019})},\ \Eprint
  {http://arxiv.org/abs/1808.08110} {arXiv:1808.08110 [hep-th]} \BibitemShut
  {NoStop}%
\bibitem [{\citenamefont {Todorov}\ and\ \citenamefont
  {Drenska}(2018)}]{Todorov:2018yvi}%
  \BibitemOpen
  \bibfield  {author} {\bibinfo {author} {\bibfnamefont {Ivan}\ \bibnamefont
  {Todorov}}\ and\ \bibinfo {author} {\bibfnamefont {Svetla}\ \bibnamefont
  {Drenska}},\ }\bibfield  {title} {\enquote {\bibinfo {title} {{Octonions,
  exceptional Jordan algebra and the role of the group $F_4$ in particle
  physics}},}\ }\href {\doibase 10.1007/s00006-018-0899-y} {\bibfield
  {journal} {\bibinfo  {journal} {Adv. Appl. Clifford Algebras}\ }\textbf
  {\bibinfo {volume} {28}},\ \bibinfo {pages} {82 arXiv:1911.13124 [hep--th]}
  (\bibinfo {year} {2018})},\ \Eprint {http://arxiv.org/abs/1805.06739}
  {arXiv:1805.06739 [hep-th]} \BibitemShut {NoStop}%
\bibitem [{\citenamefont {Todorov}(2020)}]{Todorov:2020zae}%
  \BibitemOpen
  \bibfield  {author} {\bibinfo {author} {\bibfnamefont {Ivan}\ \bibnamefont
  {Todorov}},\ }\bibfield  {title} {\enquote {\bibinfo {title} {{J}ordan
  algebra approach to finite quantum geometry},}\ }in\ \href {\doibase
  10.22323/1.376.0163} {\emph {\bibinfo {booktitle} {PoS}}},\ Vol.\ \bibinfo
  {volume} {CORFU2019}\ (\bibinfo {year} {2020})\ p.\ \bibinfo {pages}
  {163}\BibitemShut {NoStop}%
\bibitem [{\citenamefont {Ablamowicz}(1995)}]{ablamoowicz}%
  \BibitemOpen
  \bibfield  {author} {\bibinfo {author} {\bibfnamefont {Rafal}\ \bibnamefont
  {Ablamowicz}},\ }\bibfield  {title} {\enquote {\bibinfo {title} {Construction
  of spinors via {W}itt decomposition and primitive idempotents: A review},}\
  }in\ \href@noop {} {\emph {\bibinfo {booktitle} {Clifford algebras and spinor
  structures}}},\ \bibinfo {editor} {edited by\ \bibinfo {editor}
  {\bibfnamefont {Rafal}\ \bibnamefont {Ablamowicz}}\ and\ \bibinfo {editor}
  {\bibfnamefont {P.}~\bibnamefont {Lounesto}}}\ (\bibinfo  {publisher} {Kluwer
  Acad. Publ.},\ \bibinfo {year} {1995})\ p.\ \bibinfo {pages}
  {113}\BibitemShut {NoStop}%
\bibitem [{\citenamefont {Baez}(2002)}]{baez2001octonions}%
  \BibitemOpen
  \bibfield  {author} {\bibinfo {author} {\bibfnamefont {John~C.}\ \bibnamefont
  {Baez}},\ }\bibfield  {title} {\enquote {\bibinfo {title} {The octonions},}\
  }\href@noop {} {\bibfield  {journal} {\bibinfo  {journal}
  {Bull.Am.Math.Soc.}\ }\textbf {\bibinfo {volume} {39}} (\bibinfo {year}
  {2002})},\ \Eprint {http://arxiv.org/abs/math/0105155} {arXiv:math/0105155
  [math.RA]} \BibitemShut {NoStop}%
\bibitem [{\citenamefont {Baez}(2011)}]{Baez_2011}%
  \BibitemOpen
  \bibfield  {author} {\bibinfo {author} {\bibfnamefont {John~C.}\ \bibnamefont
  {Baez}},\ }\bibfield  {title} {\enquote {\bibinfo {title} {Division algebras
  and quantum theory},}\ }\href {\doibase 10.1007/s10701-011-9566-z} {\bibfield
   {journal} {\bibinfo  {journal} {Foundations of Physics}\ }\textbf {\bibinfo
  {volume} {42}},\ \bibinfo {pages} {819--855} (\bibinfo {year}
  {2011})}\BibitemShut {NoStop}%
\bibitem [{\citenamefont {Baez}\ and\ \citenamefont {Huerta}(2009
  arXiv:0904.1556 [hep-th])}]{baez2009algebra}%
  \BibitemOpen
  \bibfield  {author} {\bibinfo {author} {\bibfnamefont {John~C.}\ \bibnamefont
  {Baez}}\ and\ \bibinfo {author} {\bibfnamefont {John}\ \bibnamefont
  {Huerta}},\ }\href@noop {} {\enquote {\bibinfo {title} {The algebra of grand
  unified theories},}\ } (\bibinfo {year} {2009 arXiv:0904.1556 [hep-th]}),\
  \Eprint {http://arxiv.org/abs/0904.1556} {arXiv:0904.1556 [hep-th]}
  \BibitemShut {NoStop}%
\bibitem [{\citenamefont {Perelman}(2019)}]{Perelman}%
  \BibitemOpen
  \bibfield  {author} {\bibinfo {author} {\bibfnamefont {Carlos~Castro}\
  \bibnamefont {Perelman}},\ }\bibfield  {title} {\enquote {\bibinfo {title}
  {${R\times C\times H\times O}$ valued gravity as a grand unified field
  theory},}\ }\href@noop {} {\bibfield  {journal} {\bibinfo  {journal}
  {Advances in Applied Clifford Algebras}\ }\textbf {\bibinfo {volume} {29}},\
  \bibinfo {pages} {22} (\bibinfo {year} {2019})}\BibitemShut {NoStop}%
\bibitem [{\citenamefont {Gillard}\ and\ \citenamefont
  {Gresnigt}(2019{\natexlab{a}})}]{Gillard2019}%
  \BibitemOpen
  \bibfield  {author} {\bibinfo {author} {\bibfnamefont {Adam~B.}\ \bibnamefont
  {Gillard}}\ and\ \bibinfo {author} {\bibfnamefont {Niels~G.}\ \bibnamefont
  {Gresnigt}},\ }\bibfield  {title} {\enquote {\bibinfo {title} {Three fermion
  generations with two unbroken gauge symmetries from the complex sedenions},}\
  }\href {\doibase 10.1140/epjc/s10052-019-6967-1} {\bibfield  {journal}
  {\bibinfo  {journal} {The European Physical Journal C}\ }\textbf {\bibinfo
  {volume} {79}},\ \bibinfo {pages} {446, arXiv:1904.03186 [hep--th]} (\bibinfo
  {year} {2019}{\natexlab{a}})}\BibitemShut {NoStop}%
\bibitem [{\citenamefont {Stoica}(2018)}]{Stoica}%
  \BibitemOpen
  \bibfield  {author} {\bibinfo {author} {\bibfnamefont {Ovidiu~Cristinel}\
  \bibnamefont {Stoica}},\ }\bibfield  {title} {\enquote {\bibinfo {title} {The
  standard model algebra ({Leptons}, quarks and gauge from the complex algebra
  {Cl(6)})},}\ }\href@noop {} {\bibfield  {journal} {\bibinfo  {journal}
  {Advances in Applied Clifford Algebras}\ }\textbf {\bibinfo {volume} {28}},\
  \bibinfo {pages} {52 arXiv:1702.04336 [hep--th]} (\bibinfo {year}
  {2018})}\BibitemShut {NoStop}%
\bibitem [{\citenamefont {Yokota}(2009)}]{Yokota}%
  \BibitemOpen
  \bibfield  {author} {\bibinfo {author} {\bibfnamefont {Ichiro}\ \bibnamefont
  {Yokota}},\ }\bibfield  {title} {\enquote {\bibinfo {title} {Exceptional
  {L}ie groups},}\ }\href@noop {} {\ \textbf {\bibinfo {volume} {arXiv:0902.043
  [math.DG]}} (\bibinfo {year} {2009})}\BibitemShut {NoStop}%
\bibitem [{\citenamefont {Dray}\ and\ \citenamefont {Manogue}(1999 2901
  arXiv:math-ph/9910004v2)}]{Dray1}%
  \BibitemOpen
  \bibfield  {author} {\bibinfo {author} {\bibfnamefont {Tevian}\ \bibnamefont
  {Dray}}\ and\ \bibinfo {author} {\bibfnamefont {Corinne}\ \bibnamefont
  {Manogue}},\ }\bibfield  {title} {\enquote {\bibinfo {title} {The exceptional
  {J}ordan eigenvalue problem},}\ }\href@noop {} {\bibfield  {journal}
  {\bibinfo  {journal} {Int. J. Theo. Phys.}\ }\textbf {\bibinfo {volume} {28}}
  (\bibinfo {year} {1999 2901 arXiv:math-ph/9910004v2})}\BibitemShut {NoStop}%
\bibitem [{\citenamefont {Dray}\ and\ \citenamefont {Manogue}(2010)}]{Dray2}%
  \BibitemOpen
  \bibfield  {author} {\bibinfo {author} {\bibfnamefont {Tevian}\ \bibnamefont
  {Dray}}\ and\ \bibinfo {author} {\bibfnamefont {Corinne}\ \bibnamefont
  {Manogue}},\ }\bibfield  {title} {\enquote {\bibinfo {title} {Octonions,
  {E}$_6$ and particle physics},}\ }\href@noop {} {\bibfield  {journal}
  {\bibinfo  {journal} {J.Phys.Conf.Ser.}\ }\textbf {\bibinfo {volume} {254}},\
  \bibinfo {pages} {012005 arXiv:0911.2253} (\bibinfo {year}
  {2010})}\BibitemShut {NoStop}%
\bibitem [{\citenamefont {Lisi}(2007)}]{lisi2007exceptionally}%
  \BibitemOpen
  \bibfield  {author} {\bibinfo {author} {\bibfnamefont {A.~Garrett}\
  \bibnamefont {Lisi}},\ }\bibfield  {title} {\enquote {\bibinfo {title} {An
  exceptionally simple theory of everything},}\ }\href@noop {} {\ \textbf
  {\bibinfo {volume} {arXiv:0711.0770 [hep-th]}} (\bibinfo {year}
  {2007})}\BibitemShut {NoStop}%
\bibitem [{\citenamefont {Furey}(2016)}]{Fureycharge}%
  \BibitemOpen
  \bibfield  {author} {\bibinfo {author} {\bibfnamefont {Cohl}\ \bibnamefont
  {Furey}},\ }\bibfield  {title} {\enquote {\bibinfo {title} {Charge
  quantisation from a number operator},}\ }\href@noop {} {\bibfield  {journal}
  {\bibinfo  {journal} {Phys. Lett. B}\ }\textbf {\bibinfo {volume} {742}},\
  \bibinfo {pages} {195 DOI:10.1016/j.physletb.2015.01.023 r arXiv:1603.04078v2
  [hep--th]} (\bibinfo {year} {2016})}\BibitemShut {NoStop}%
\bibitem [{\citenamefont {Gillard}\ and\ \citenamefont
  {Gresnigt}(2019{\natexlab{b}})}]{Gillard}%
  \BibitemOpen
  \bibfield  {author} {\bibinfo {author} {\bibfnamefont {Adam~B.}\ \bibnamefont
  {Gillard}}\ and\ \bibinfo {author} {\bibfnamefont {Niels}\ \bibnamefont
  {Gresnigt}},\ }\bibfield  {title} {\enquote {\bibinfo {title} {The {Cl(8)}
  algebra of three fermion generations with spin and full internal
  symmetries},}\ }\href@noop {} {\ \textbf {\bibinfo {volume}
  {arXiv:1906.05102}} (\bibinfo {year} {2019}{\natexlab{b}})}\BibitemShut
  {NoStop}%
\bibitem [{\citenamefont {Vivan~Bhatt}\ and\ \citenamefont {Singh}(2021 to be
  published in J. Phys. G: Nuclear and Particle Physics
  https://doi.org/10.1088/1361-6471/ac4c91)}]{vvs}%
  \BibitemOpen
  \bibfield  {author} {\bibinfo {author} {\bibfnamefont {Vatsalya~Vaibhav}\
  \bibnamefont {Vivan~Bhatt}, \bibfnamefont {Rajrupa~Mondal}}\ and\ \bibinfo
  {author} {\bibfnamefont {Tejinder~P.}\ \bibnamefont {Singh}},\ }\bibfield
  {title} {\enquote {\bibinfo {title} {Majorana neutrinos, exceptional {J}ordan
  algebra, and mass ratios of charged fermions},}\ }\href@noop {} {\bibfield
  {journal} {\bibinfo  {journal} {arXiv:2108.05787v2 [hep-ph]}\ } (\bibinfo
  {year} {2021 to be published in J. Phys. G: Nuclear and Particle Physics
  https://doi.org/10.1088/1361-6471/ac4c91})}\BibitemShut {NoStop}%
\bibitem [{\citenamefont {Vaibhav}\ and\ \citenamefont
  {Singh}(2021)}]{Vatsalya1}%
  \BibitemOpen
  \bibfield  {author} {\bibinfo {author} {\bibfnamefont {Vatsalya}\
  \bibnamefont {Vaibhav}}\ and\ \bibinfo {author} {\bibfnamefont {Tejinder~P}\
  \bibnamefont {Singh}},\ }\bibfield  {title} {\enquote {\bibinfo {title}
  {Left-right symmetric fermions and sterile neutrinos from complex split
  biquaternions and bioctonions},}\ }\href
  {https://arxiv.org/pdf/2108.01858.pdf} {\bibfield  {journal} {\bibinfo
  {journal} {arXiv:2108.01858 [hep-ph]}\ } (\bibinfo {year}
  {2021})}\BibitemShut {NoStop}%
\bibitem [{\citenamefont {Adler}(2004)}]{Adler:04}%
  \BibitemOpen
  \bibfield  {author} {\bibinfo {author} {\bibfnamefont {Stephen~L.}\
  \bibnamefont {Adler}},\ }\href@noop {} {\emph {\bibinfo {title} {Quantum
  theory as an emergent phenomenon}}}\ (\bibinfo  {publisher} {Cambridge
  University Press},\ \bibinfo {year} {2004})\BibitemShut {NoStop}%
\bibitem [{\citenamefont {Singh}(2020{\natexlab{a}})}]{Singh2020DA}%
  \BibitemOpen
  \bibfield  {author} {\bibinfo {author} {\bibfnamefont {Tejinder~P.}\
  \bibnamefont {Singh}},\ }\bibfield  {title} {\enquote {\bibinfo {title}
  {Trace dynamics and division algebras: towards quantum gravity and
  unification.}}\ }\href@noop {} {\bibfield  {journal} {\bibinfo  {journal}
  {Zeitschrift f\"ur Naturforschung A}\ }\textbf {\bibinfo {volume} {76}},\
  \bibinfo {pages} {131, DOI: https://doi.org/10.1515/zna--2020--0255,
  arXiv:2009.05574v44 [hep--th]} (\bibinfo {year}
  {2020}{\natexlab{a}})}\BibitemShut {NoStop}%
\bibitem [{\citenamefont {Chamseddine}\ and\ \citenamefont
  {Connes}(1997)}]{Chams:1997}%
  \BibitemOpen
  \bibfield  {author} {\bibinfo {author} {\bibfnamefont {Ali~H.}\ \bibnamefont
  {Chamseddine}}\ and\ \bibinfo {author} {\bibfnamefont {Alain}\ \bibnamefont
  {Connes}},\ }\bibfield  {title} {\enquote {\bibinfo {title} {The spectral
  action principle},}\ }\href@noop {} {\bibfield  {journal} {\bibinfo
  {journal} {Commun. Math. Phys.}\ }\textbf {\bibinfo {volume} {186}},\
  \bibinfo {pages} {731 arXiv:hep--th/9606001} (\bibinfo {year}
  {1997})}\BibitemShut {NoStop}%
\bibitem [{\citenamefont {Landi}\ and\ \citenamefont
  {Rovelli}(1997)}]{Rovelli}%
  \BibitemOpen
  \bibfield  {author} {\bibinfo {author} {\bibfnamefont {Giovanni}\
  \bibnamefont {Landi}}\ and\ \bibinfo {author} {\bibfnamefont {Carlo}\
  \bibnamefont {Rovelli}},\ }\bibfield  {title} {\enquote {\bibinfo {title}
  {General relativity in terms of {Dirac} eigenvalues},}\ }\href@noop {}
  {\bibfield  {journal} {\bibinfo  {journal} {Phys. Rev. Lett.}\ }\textbf
  {\bibinfo {volume} {78}},\ \bibinfo {pages} {3051 arXiv:gr--qc/9612034}
  (\bibinfo {year} {1997})}\BibitemShut {NoStop}%
\bibitem [{\citenamefont {Singh}(2020{\natexlab{b}})}]{Singhspin1}%
  \BibitemOpen
  \bibfield  {author} {\bibinfo {author} {\bibfnamefont {Tejinder~P.}\
  \bibnamefont {Singh}},\ }\bibfield  {title} {\enquote {\bibinfo {title} {A
  basic definition of spin in the new matrix dynamics},}\ }\href@noop {}
  {\bibfield  {journal} {\bibinfo  {journal} {Zeitschrift f\"ur Naturforschung
  A}\ }\textbf {\bibinfo {volume} {75}},\ \bibinfo {pages} {963,
  arXiv:2006.16274v1} (\bibinfo {year} {2020}{\natexlab{b}})}\BibitemShut
  {NoStop}%
\end{thebibliography}%

\end{document}